\documentstyle[12pt,epsfig,psfig]{article}
\newlength{\capwidth}
\newcommand{\BC}{\begin{center}}
\newcommand{\EC}{\end{center}}
\newcommand{\ALFAS}{$\alpha_{s}$ }

\newcommand{\mrm}{\mathrm}

\renewcommand{\bigskip  }{\vskip 0.2in}

\def\Zff{\mathrm{ Z \rightarrow f\bar{f}}}
\def\Zqq{\mathrm{ Z \rightarrow hadrons }}
\def\Zbb{\mathrm{ Z \rightarrow b\bar{b}}}

\def\DR {\Delta\rho}

\def\dv{\mathrm{\delta_{vtb}}}

\def\SWA{\sin^2\theta_{eff}^{lept}}

\def\CW1{\cos\theta_W}
\def\FR{\begin{flushright}}
\def\FL{\begin{flushleft}}
\def\EFR{\end{flushright}}
\def\EFL{\end{flushleft}}

\def\Bb0{\mathrm{\bar{B^o}}}
\def\B0{\mathrm{B^o}}

\begin{document}
%%%%%%%%%%%%%%%%%%%%%%%%%%%%%%%%%%%%%%   Start some addition %%%%%%%%%%%%
%%%%%%%%%%%%%%%%%%% Title %%%%%%%%%%%%%%%%%%
\begin{center}
\Large
EUROPEAN ORGANIZATION FOR NUCLEAR RESEARCH
\end{center}

\vspace*{3mm}
\begin{flushright}
% \large
CERN-OPEN-2001-067 

\vspace*{1mm}
\noindent
August 29, 2001
\end{flushright}

\vspace{1.5cm}
\begin{center}
{\Large\bf Precision testing the Standard Model}
%%%%%%%%%%%%%%%%%%% Abstracts %%%%%%%%%%%%%%%%

\vspace*{1.5cm}
{\large T. Aziz$^{\star,~a}$ and A. Gurtu$^{\star,~b}$}

EP Division, CERN, 1211 Geneva 23, Switzerland

\vspace*{1.5cm}
{\Large\em Abstract}
\end{center}
\noindent
Internal consistency of the most sensitive electroweak measurements within the
standard model framework is examined. Confirming an earlier observation on 
the separation of Z-pole asymmetry measurements into {\em hadronisation-free}
and {\em hadronisation-sensitive}, the electroweak mixing angle derived 
using the former is in perfect agreement with the precision W mass. 
These two complimentary measurements of weak radiative corrections, when 
combined with the lower limit on Higgs mass, are incompatible with
the measured top quark mass.  To overcome this inconsistency, a
scenario readily testable in Run-II at Tevatron is envisaged: an upward shift of the 
top quark mass by about 10 GeV ($\sim 2\sigma$). If, however, the improved top 
quark mass remains at its current value or the lower limit on Higgs mass moves up 
substantially, then abandoning the SM may become inevitable.

\vfill
\noindent
$^{\star}$On sabbatical leave from Tata Institute of Fundamental Research,
Mumbai, India. \\
$^{a}$ e-mail: Tariq.Aziz@cern.ch \\
$^{b}$ e-mail: Atul.Gurtu@cern.ch

\begin{figure}[tb] 
\begin{center}  
\vspace*{-1.0cm}
\mbox{\epsfig{file=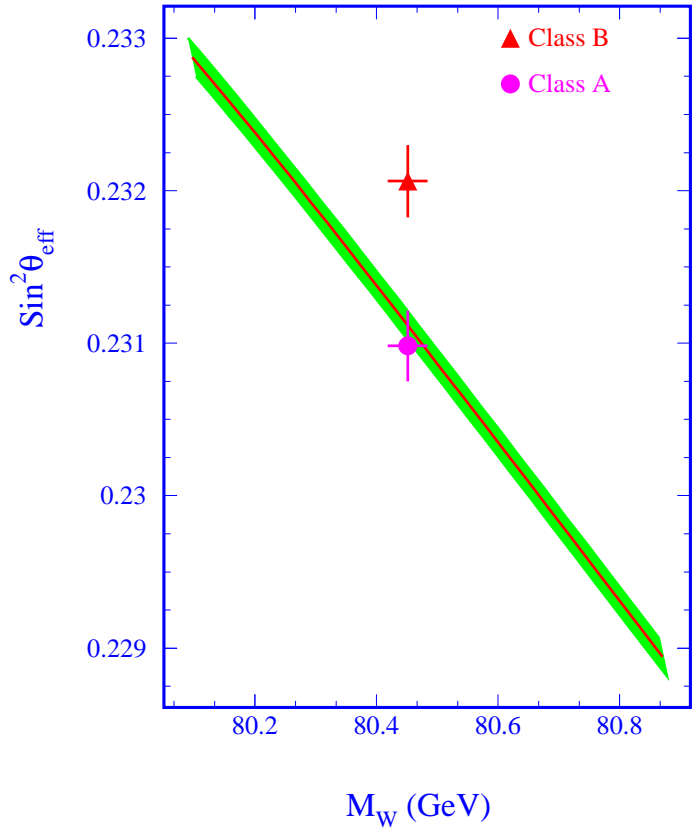,height=14cm }}
       \parbox{\capwidth}{
\vspace*{-1.50cm}
\caption{\small Dependence of $\mrm\SWA$\ on $M_W$.
The band around central line 
corresponds to uncertainty in $\alpha(M_Z)$}
  \label{fig1:seffw}
}
   \end{center}
 \end{figure}

\begin{figure}[t] 
\begin{center}  
%    \mbox{\epsfig{file=ta3.ps,width=5cm}%
%\hspace*{2.0cm}\epsfig{file=ta5.ps,width=5cm}}
\hspace*{-1.0cm}
\mbox{\epsfig{file=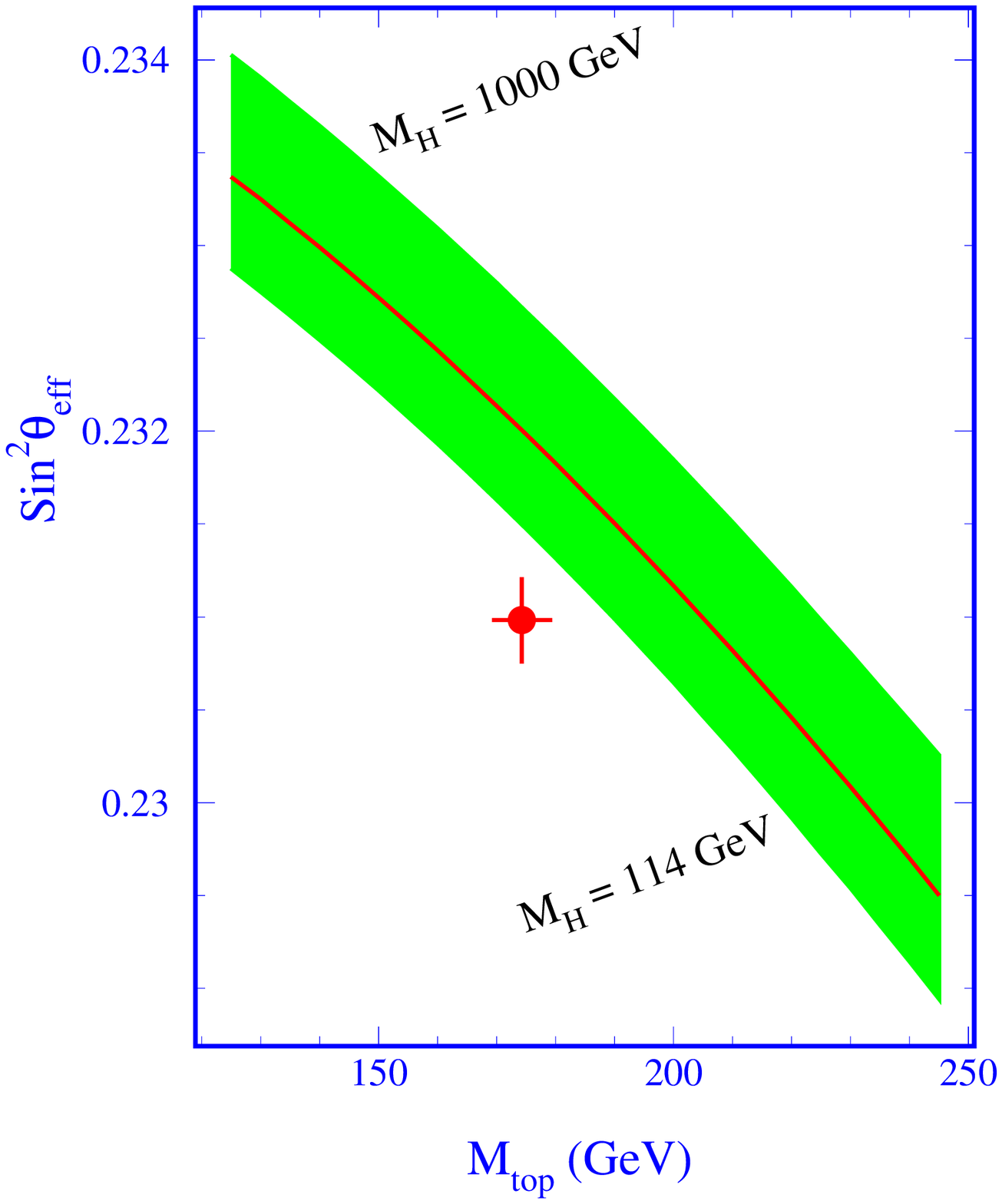, height=12cm }%
\hspace*{-1.5cm}\epsfig{file=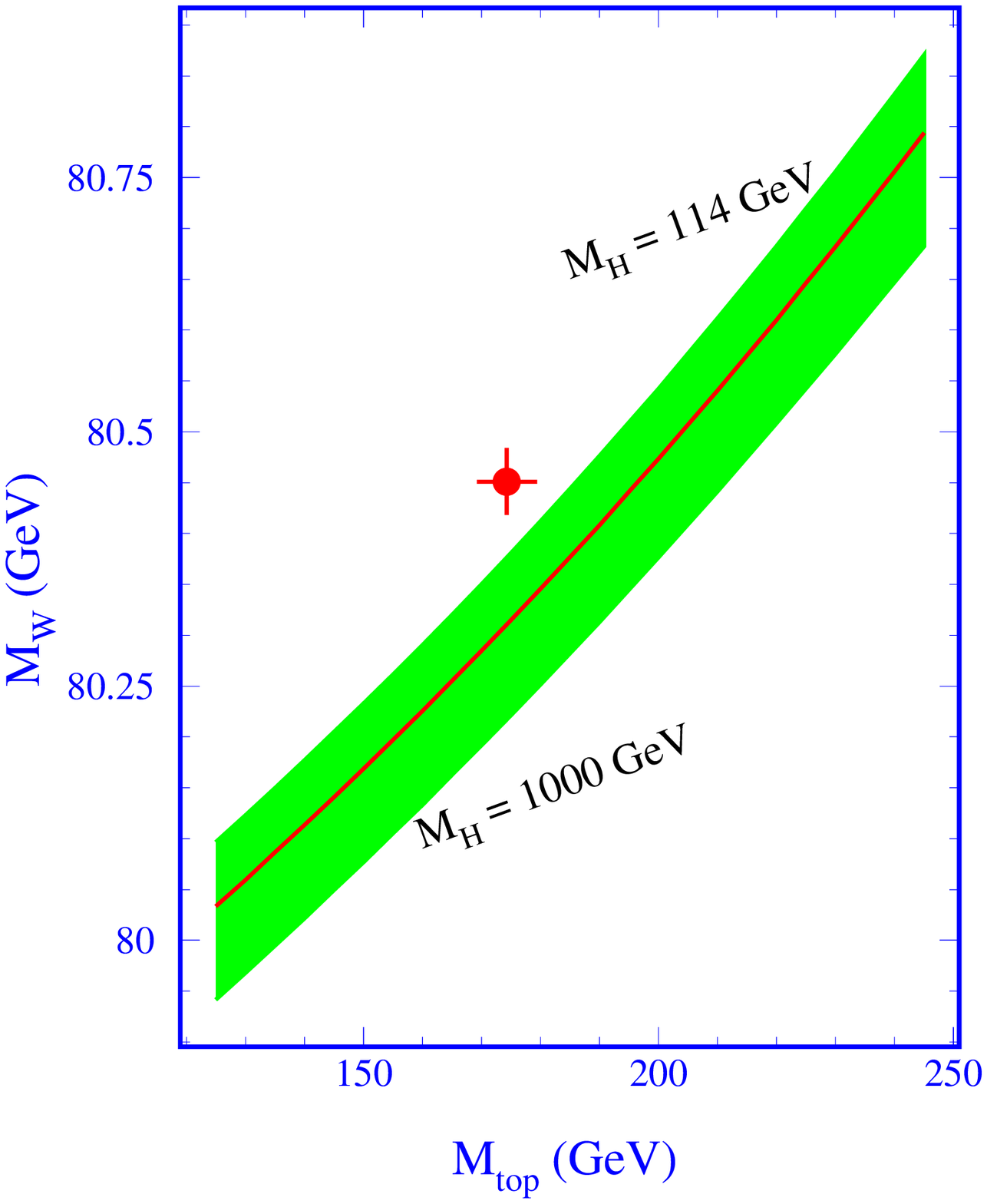, height=12cm}}
       \parbox{\capwidth}{
\vspace*{-1.50cm}
\caption{\small Dependence of $\mrm\SWA$\ on $M_{top}$ (left) and $M_W$
on $M_{top}$ (right). The band around central line in both 
corresponds to  Higgs dependence. The edges corresponding to  Higgs mass
of 114 GeV and  1000 GeV are indicated. The central line corresponds to a Higgs mass
of 300 GeV in both.}
  \label{fig2:sefft}
}
   \end{center}
 \end{figure}

\begin{figure}[t] 
\begin{center}  
\mbox{\epsfig{file=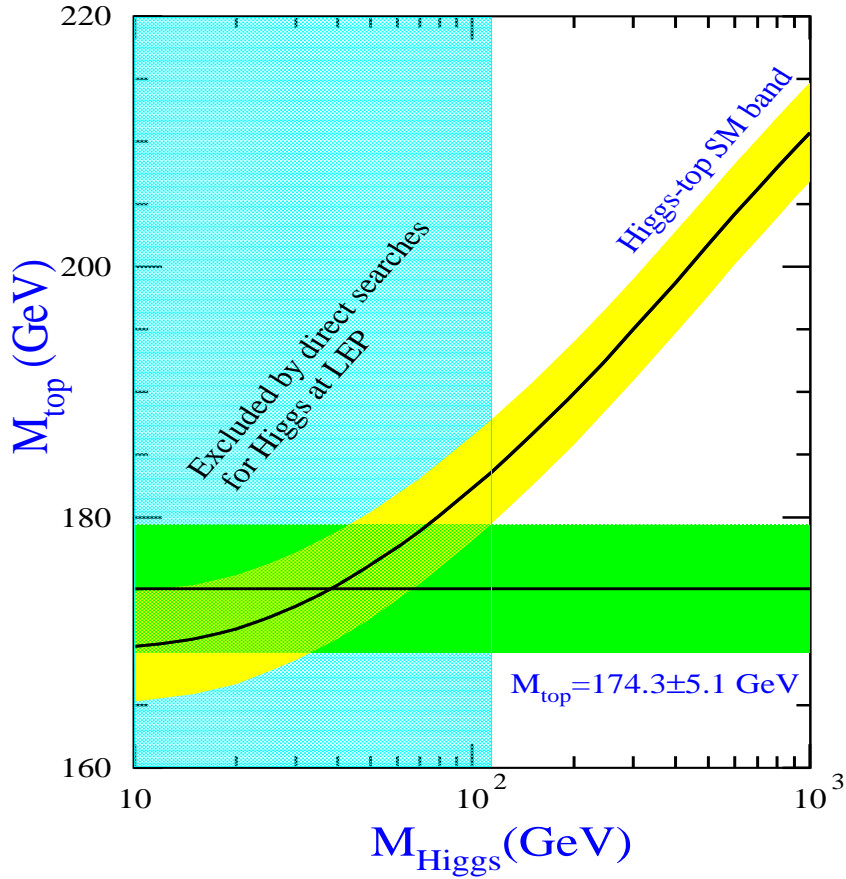,height=15cm, width=15cm }}
       \parbox{\capwidth}{
\vspace*{-1.50cm}
\caption{\small $M_{top}$ as function of Higgs mass for which fit to data
give minimum  $\chi^2$. The band {\em Higgs-top SM band} corresponds
to $1\sigma$ fit uncertainty and the central line as the best value.
The measured top mass (174.3$\pm$5.1 GeV) is depicted as a band parallel to Higgs axis. 
The Higgs mass excluded region below 114.1 GeV is also indicated.}

  \label{fig3:mtmh}
}
   \end{center}
 \end{figure}

\newpage
\noindent 
Over the last several years the Standard Model (SM) of electroweak interactions 
has been subjected to precision tests through a variety  measurements 
at LEP, SLC and Tevatron. While the measurements at Z pole are sensitive to the
presence of top quark and Higgs boson via weak radiative corrections, the direct
measurement of top quark mass at Tevatron and W boson mass at LEP2 and Tevatron,
along with the lower bound on Higgs mass at LEP2, provide a detailed set 
to test the validity of SM. These measurements, which are now final or almost final, have 
reached a precision where it is possible to verify their self-consistency within the SM 
framework. In case of serious inconsistency either this framework collapses or the data 
have to be re-looked. In this paper we examine the self-consistency of these data and 
discuss their implications. \\
\\
The important weak radiative corrections that affect $\Zff$ (f: fermion) widths 
and asymmetries~\cite{yellowrep} appear as propagator correction, $\DR$, sensitive to
both top and Higgs masses, and $\Zbb$ width where there is an additional vertex 
correction, $\dv$, sensitive only to top quark mass. All those measurements where ratio of
widths are involved, like  $R_\ell$ and $\sigma^{peak}$, the $\DR$ sensitivity is 
practically lost and whatever top dependence one sees is due to indirect effect of 
$\dv$ in $\Zbb$ width. Thus the only sensitive measurements to test $\DR$ are fermion 
asymmetries leading to effective weak mixing angle\footnote{In terms of vector and axial 
vector couplings $g_{Vf}, g_{Af}$, the definition of the effective weak mixing angle is
$\mrm\SWA  =  \frac{1}{4}(1 - g_V/g_A)$},
%%%%%%%%%%%%%%%%%%%%
$\mrm\SWA$, and the total Z decay width, $\Gamma_Z$. 
The ratio $R_b =({{\Zbb}/{\Zqq}})$  is a clean measure of $\dv$ and practically
independent of the Higgs mass. We will restrict ourselves to only those measurements 
whose sensitivity for weak radiative effects is well above the measurement uncertainties. 
On the Z pole these are $\mrm\SWA$, $R_b$ and $\Gamma_z.$ The first two are free from 
\ALFAS  uncertainty, the last one requires \ALFAS for its interpretation.\\        
\\
Observations on the measurements of asymmetries and hence $\mrm\SWA$ over the years  
have shown a clear pattern and several years ago it was pointed out~\cite{swhad} that all 
asymmetry measurements should be placed in two classes. Class A measurements  
where {\em hadronisation effects} are not relevant for the final result and 
class B measurements where {\em hadronisation effects} cannot be avoided 
and can only be corrected with whatever understanding of these phenomena we 
have. In class A measurements are forward-backward asymmetry of leptons $e, \mu, \tau$ at
LEP and left-right asymmetry at SLC. In class B are measurements of all quark asymmetries
as well as $\tau$ polarisation asymmetry measured through hadronic decays. 
The two class of measurements were found to be more than $3\ \sigma$ apart already in 
1997~\cite{swhad}. It was also pointed out that the measured W mass was in favour of
class A measurements, though W mass was much less precise than now. With increased
precision of the W mass with time, its consistency with class A asymmetry measurements 
and incompatibility with class B measurements has become even more pronounced. \\
\\
Meanwhile the lower limit on Higgs mass from direct searches at LEP has increased 
considerably: it is now 114.1 GeV~\cite{mhg1} (with even a $\sim 2\sigma$ hint of
Higgs at $\sim$115 GeV). This increase in lower bound on Higgs mass along with the
improvement in W mass helps significantly in testing the self-consistency of data. 
The latest compilation of all the electroweak measurements from LEP, SLC and
Tevatron is taken from ~\cite{mwg1}. With the above mentioned classification of
asymmetry data, the average $\mrm\SWA$ for the two classes of measurements from LEP and 
SLC are: $0.23098 \pm 0.00023$ for class A ({\em hadronisation-free}: SLD-dominated) and 
$0.23206 \pm 0.00024$ for class B ({\em hadronisation-sensitive}: LEP-dominated) 
and these are $3.3 \sigma$ apart.
 The measured  W mass from LEP and Tevatron has 
improved considerably and the current  average value is $80.451\pm 0.033$ GeV. The top 
mass measured at Tevatron is $174.3 \pm 5.1$ GeV and the lower bound on 
Higgs boson mass is 114.1 GeV at 95\% CL from LEP2 searches.\\ 
% In addition the measured 
% Z decay width, $\Gamma_Z = 2.4952\pm 0.0023$ GeV and $R_b = 0.21646\pm 0.00065$\\
\\
In parenthesis we  mention that this problem is being acknowledged  
by the LEP experiments~\cite{mwg1} as a discrepancy between $\mrm\SWA$
determined using {\em leptonic} asymmetries and {\em hadronic} asymmetries. The
average  values for the two types  being $0.23113\pm 0.00021$
and $0.23230\pm 0.00029$ respectively and the two differ by $3.3 \sigma$.\footnote{ 
The discrepancy is mainly driven by two most precise measurements. The
b quark forward-backward asymmetry, $\mrm A_{fb}^b$, at LEP  and the left-right
asymmetry, $\mrm A_{LR}$, at SLC. Since $\mrm A_{fb}^b$ is 
% To an excellent approximation
% $\mrm A_{fb}^b = {3\over 4}{A_b\ A_e} \cong 1.395(1 - 4 \SWA)$,
pretty insensitive to vertex corrections, any new physics explanation is unlikely
to resolve such large discrepancy. More practical
approach would be to treat the discrepancy due to either systematic problem~\cite{swhad} or 
strong statistical fluctuation~\cite{mwg1}.}\\   
%%%
\\ 
The simplest consistency test of data is provided by comparison of $\mrm \SWA $ 
measurements with the W boson mass, as both are affected by weak radiative effects 
arising from top and Higgs and provide a complementary measure of the same weak 
corrections. These measurements are  compared in figure~\ref{fig1:seffw}.
%%%%%%%%%%%%%%%%%%%%%%%%%%%%%%%%%%%%%%%%%%%%%%%%%%%%%
%%%%%%%%%%%%%%%%%%%%%%%%%%%%%%%%%%%%%%%%%%%%%%%%%%%%%%%%%%%%%%%%%%%%%%%%%%%%%%%%%%%%
This figure clearly illustrates the  difference between the Class A and 
Class B measurements of $\mrm\SWA$\ as well as a strong preference for Class A due to
pretty precise W mass.
Thus, as already emphasised earlier~\cite{swhad}, for further consideration we prefer 
$\mrm\SWA$ from class A measurements only and suggest that till a better understanding 
of class B measurements is found, they be kept aside.\\
\\
As the next step we study the consistency of  $\mrm\SWA$ or $M_W$  with that of the
measured top quark mass. This is shown in figure~\ref{fig2:sefft}. One sees explicitly
the effect of the recent LEP lower limit on the Higgs mass: it is clear that for 
$M_{Higgs}> 114.1$\ GeV the present value of top quark mass is not sufficient to drive 
the radiative corrections to explain either $\mrm\SWA$\ or $M_W$. It seems impossible to
satisfy all the measurement constraints, $\mrm\SWA$,  $M_W$, $M_{top}$\ and 
$M_{Higgs}> 114.1$\ GeV 
simultaneously within the SM framework. One has to make a choice: either one goes beyond
the scope of the SM and formulates a model in which all the existing data can be
explained, or one can examine the data critically and see if a reasonable change in one
of the measurements would remove the discrepancy within the SM. There have been
recent approaches using the first option~\cite{alta01}. In this paper we explore the
second alternative. We note, firstly, that two precise and independent measurements, 
$\mrm\SWA$\ and $M_W$, are self-consistent within the SM framework.
Secondly, in relative terms $M_{top}$\ is presently the least accurately determined data
point, being determined to an accuracy of 2.9\%, whereas $\mrm\SWA$\ and $M_W$\ are
determined to 0.1\% and 0.04\% respectively. Thus the minimal change in data which may
make it consistent within the SM is an upward movement of $M_{top}$. We now try to 
estimate how much change in it is required for everything to fit together.\\
%%%%%%%%%%%%%%%%%%%%%%%%%%%%%%%%%%%%%%%%%%%%%%%%%%%%%%%%%%%%%%%%%%%%%%%%%%%%%%%%%%%%
%%% \noindent
\\
To arrive at a quantitative estimate of required increase in $M_{top}$ to explain the data,
we have taken the most sensitive and clean measurements that are indicative of weak 
radiative effects. These are the $\mrm\SWA = 0.23098\pm 0.00023$ from 
class A ({\em hadronisation-free}) measurements,
$R_b = 0.21646\pm 0.00065$ and $\Gamma_Z = 2.4952\pm 0.0023$ GeV at the Z pole, 
$M_W = 80.451 \pm 0.033$ GeV and the lower bound of 114.1 GeV 
on the Higgs mass. All in association with the $M_Z = 91.1875\pm 0.0021$ GeV, 
$\alpha(M_Z) = 1/(128.945 \pm 0.052)$\footnote{ This corresponds to the best measured value
of $\Delta\alpha^{(5)}_{had}(M_Z) = 0.02761\pm 0.00036$ as determined using BES 
collaboration data~\cite{ghmoriond01, bpfeb01}.}       
and $\alpha_s(M_Z) = 0.1181\pm 0.0020$~\cite{pdg} as inputs and ZFITTER (version 6.36)~\cite{zfitr} as the SM package. A fit to these data is performed to extract top quark mass for 
various Higgs masses in an extended range, from 10 -- 1000 GeV. Fit results for a few
representative values of the Higgs mass are given in table~\ref{tab:mhmt} and 
in figure~\ref{fig3:mtmh} the variation of the top quark mass is shown as a function of the Higgs mass for which the  
%%%%%%%%%%%%%%%%%%%%%%%%%%%%%%%%%%%%%%%%%%%%%%%%%%%%%
%%%%%%%%%%%%%%%%%%%%%%%%%%%%%%%%%%%%%%%%%%%%%%%%%%%%%%%%%%%%%%%%%%%%%%%%%%%%%%%%%%%%
$\chi^2$ is minimum. The band indicated as the {\em Higgs-top SM band} along with 
the central line in figure~\ref{fig3:mtmh}  indicates the best line with $1\sigma$ 
fit error where weak radiative effects balance each-other
to get the minimum $\chi^2$. The band with central line parallel to Higgs-axis 
indicates the measured  top quark mass with its errors. The broad band
parallel to top-axis is the lower bound on Higgs from LEP2 searches~\cite{mhg1}. 
Given the Higgs-bound of 114.1 GeV, the only reasonable option to gain self-consistency 
of data within the SM is to move top quark mass up by about 10 GeV to $\sim$184 GeV. This
represents a 2$\sigma$\ shift upward from the current value of 174.3$\pm$5.1 GeV, which
does not seem dramatic. However, this would suffice only if the Higgs is indeed around the
corner. In case the lower bound on the Higgs mass increases substantially beyond the
current value, the change in $M_{top}$\ required for SM consistency of data would be much
more and would become increasingly unrealistic. In such a scenario the SM would really
have to be abandoned. Indeed, the variation of $\chi^2$\ with Higgs mass in 
table~\ref{tab:mhmt} indicates that data favour a light Higgs.\\
%%%%%%%%%%%%%%%%%%%%%%%%%%%%%%%%%%%%%%%%%%%%%%%%%%%%%%%%%%%%%%%%%%%%%%%%%%%%
\begin{table}[tb]
\begin{center}
\parbox{\capwidth}{\caption {Result of the fit using data most sensitive to radiative
corrections. The Higgs mass has been varied starting from the lower bound
and top mass has been extracted.}
\label{tab:mhmt} }\\
\vspace*{0.5cm}
\begin{tabular}
  {|l|c|c|c|c|c|}\hline             
Fit Parameter & \multicolumn{5}{c|}{$M_{Higgs}$(GeV)} \\
\cline{2-6}
{   }    & 114.1  &   200 &  300  & 450   &  1000 \\
\hline                                                                                 
$M_{top}$(GeV) &  $ 184\pm 4$ & $ 190\pm 4$ & $ 195\pm 4 $ &$ 200\pm 4 $ & $ 211\pm 4 $\\
\hline
$\chi^2 /DOF $ & $ 4.3/4 $ & $  6.0/4 $ & $ 7.5 $ & $ 9.2/4 $ & $13.8/4 $\\
$(Prob\ \%) $  & $ 36.7 $ & $  20.0  $ & $ 11.4 $ & $ 5.5 $ & $0.8 $\\
\hline
\end{tabular}
\end{center}
%%%%%%%%%%
\end{table}
%%%%%%%%%%%%%%%%%%%%%%%%%%%%%%%%%%%%%%%%%%%%%%%%%%%%%%%%%%%%%%%%%%%%%%%%%%%%%%%%%%%%
\\
Run-II at the Tevatron will play a crucial role in clarifying the situation: the error on
the top mass is expected to be reduced considerably and on the W mass to some extent. The
lower bound on Higgs mass may also be improved or, if nature is kind, it may be
discovered. \\
\\
We conclude that the present measurements on W mass, $\mrm\SWA$ and lower Higgs-bound 
are not compatible with the present value of the top quark mass within the standard model 
framework, the incompatibility being driven by the recent LEP2 result on the lower 
Higgs-bound. The simplest way to restore compatibility, without resorting to new physics, 
would be to move the top quark mass up from its present value of 174 GeV 
to a higher value of 184 GeV or so. If the Higgs lower bound does not move very much 
upward, this required increase of about 10 GeV is essentially a $2\sigma$ increase and 
can be tested soon in Run-II at the Tevatron. In case the top quark mass remains 
at its present value and the measurement uncertainty goes down to 3 GeV or better, or
if the lower limit on the Higgs mass increases substantially, that will be the real 
{\em beginning of the end} of the standard model.\\   
%%%%%%%%%%%%%%%%%%%%%%%%%%%%%%%%%%%%%%%%%%%%%%%%%%%%%%%%%%%%%%%%%%%%%%%%%%%%%%%%%%%%%%%%%%%%%%%


\begin{thebibliography}{199}
\bibitem{yellowrep}
M. Consoli, W. Hollik and F. Jegerlehner in: Z Physics at LEP 1, eds.
G. Altarelli, R. Kleiss and C. Verzegnassi, CERN 89-08 (1989), vol. 1, p 7. 
\bibitem{swhad}
T. Aziz, Mod. Phys. Lett. A12 (1997) 2535\\
Indications of such pattern were noticed even earlier:\\ 
ibid: Mod. Phys. Lett. A9 (1994) 1857.
\bibitem{mhg1}
ALEPH, DELPHI, L3 and OPAL Collaborations and the LEP working group for 
Higgs Boson searches, CERN preprint CERN-EP/2001-055 (2001).
\bibitem{mwg1}
Results and description from Winter 2001: \\
The LEP Collaborations ALEPH, DELPHI, L3, OPAL, the LEP Electroweak 
Working Group and the SLD Heavy Flavour and Electroweak Groups,
LEPEWWG/2001-01, 31 May 2001\\ \\
Results from Summer 2001: \\
S. Mele, Combined fit to Electroweak data and constraints on the Standard Model, talk at Intern. Europhysics Conf. on High Energy Physics, Budapest, 12-18 July 2001, \\
J. Drees, Review of Final LEP Results or A Tribute to LEP, talk at Lepton
Photon conference, LP01, Rome, 23-28 July 2001.\\
M.W. Gruenewald, `Electroweak Analyses', talk at the LEP Physics Jamboree, CERN,
10 July 2001.
\bibitem{alta01}
some recent examples: \\
M. Chanowitz, hep-ph/0104024, LBNL-47684, 2 April 2001 (revised 27 May 2001).\\
G. Altarelli et al, hep-ph/0106029, CERN preprint CERN-TH/2001-145, (2001).
\bibitem{ghmoriond01}
G. Huang, Measurement of R in 2-5 GeV with BES II at BEPC, talk at XXXVIth
Rencontres de Moriond, Electroweak Interactions and Unified Theories, Les Arcs,
France, 10-17 March 2001.
\bibitem{bpfeb01}
H. Burkhardt and B. Pietrzyk, Phys. Lett. B 513 (2001) 46.
\bibitem{pdg}
PDG Collab., D.E. Groom et al. Euro. Phys. Jour. C 15 (2000) 1.
\bibitem{zfitr}
D. Bardin et al, Z. Phys. C 44 (1989) 493; Comp. Phys. Comm. 59 (1990) 303; 
ZFITTER v.6.21: DESY 99-070 (1999) [hep-ph/9908433].
\end{thebibliography}
\end{document}